\documentclass{svproc}
\usepackage{url}
\usepackage{graphicx}
\usepackage{multicol}
\usepackage{footmisc}

\usepackage{latexsym}
\usepackage{amsmath,amssymb,graphicx,color,bm}
\usepackage{enumerate}
 \renewcommand{\And}{\wedge}
  \newcommand{\beq}{\begin{equation}}
  \newcommand{\eeq}{\end{equation}}
  \newcommand{\beql}[1]{\begin{equation}\label{eq:#1}}
  \newcommand{\beqas}{\begin{eqnarray*}}
  \newcommand{\eeqas}{\end{eqnarray*}}
\newtheorem{Proposition}[theorem]{Proposition}
 \newcommand{\bTheorem}{\begin{theorem}}
\newcommand{\eTheorem}{\end{theorem}}
\newcommand{\bProof}{\begin{proof}}
\newcommand{\eProof}{\end{proof}}

  \newcommand{\al}{\alpha}
  \newcommand{\be}{\beta}
  \newcommand{\mb}{\mbox}
  
  \newcommand{\ph}{\phi}
 \newcommand{\ps}{\psi}
  \newcommand{\De}{\Delta}
  \newcommand{\Eq}[1]{Eq.~(\ref{eq:#1})}
  \newcommand{\Iff}{\Leftrightarrow}
  \newcommand{\Inf}{\bigwedge}
  \newcommand{\Not}{\neg}
  \newcommand{\Or}{\vee}
  \newcommand{\Sup}{\bigvee}
  \newcommand{\THEN}{\Rightarrow}
\newcommand{\Then}{\rightarrow}
\newcommand{\Thenj}{\rightarrow}
\newcommand{\Implies}{\rightarrow}
\renewcommand{\Iff}{\leftrightarrow}
\newcommand{\V}{V}
  \DeclareMathOperator{\dom}{dom}
\renewcommand{\iff}{\quad\mbox{iff}\quad}

\newcommand{\bracket}[1]{\langle#1\rangle}
\newcommand{\av}{\bracket}
  \newcommand{\rank}{\mbox{\rm rank}}
 \newcommand{\cA}{{\cal A}}
  \newcommand{\cB}{{\cal B}}
  \newcommand{\cC}{{\cal C}}
  \newcommand{\cD}{{\dom}}
  \newcommand{\cF}{{\cal F}}
  \newcommand{\cH}{{\cal H}}
  \newcommand{\cL}{{\cal L}}
  \newcommand{\cM}{{\cal M}}
  \newcommand{\cP}{{\cal P}}
  \newcommand{\cQ}{{\cal Q}}
  \newcommand{\cR}{{\cal R}}

\newcommand{\VQH}{\V^{(\cQ(\cH))}}
\newcommand{\VQ}{\V^{(\cQ)}}
\newcommand{\VL}{\V^{(\cQ)}}
\newcommand{\VB}{\V^{(\cB)}}

\newcommand{\LL}{\cL}

\renewcommand{\inf}{\bigwedge}
\renewcommand{\sup}{\bigvee}
\newcommand{\val}[1]{[\![#1]\!]}
\newcommand{\vval}[1]{[\![#1]\!]_{\cQ}}
\newcommand{\vvall}[1]{[\![#1]\!]_{\cQ(\cH)}}
\newcommand{\valj}[1]{[\![#1]\!]}
\newcommand{\tcom}{\,\rotatebox[origin=c]{90}{$\models$}\,}
\newcommand{\commutes}{\,\rotatebox[origin=c]{270}{$\multimap$}\,}
\newcommand{\cuniv}{\underline{\Or}}
\newcommand{\cm}{\underline{\Or}}
\newcommand{\com}{{\tcom}}
\newcommand{\p}{{}^{\perp}}
\newcommand{\ck}[1]{\check{#1}}
\newcommand{\id}{{\rm id}}
\newcommand{\benum}{\begin{enumerate}[{\rm (i)}]\itemsep=0in}
\newcommand{\eenum}{\end{enumerate}}

\newcommand{\ie}{{\it i.e.}}
\newcommand{\BH}{\cB(\cH)}

\begin{document}
\mainmatter    
\title{Reforming Takeuti's Quantum Set Theory\\ to Satisfy 
De Morgan's Laws}
\titlerunning{Quantum Set Theory}  
\author{\sc Masanao Ozawa\inst{1}\inst{2}}
\authorrunning{Masanao Ozawa} 
\institute{{\it Graduate School of Informatics,
Nagoya University, Chikusa-ku, Nagoya, 464-8601, Japan}\\
\email{ozawa@is.nagoya-u.ac.jp}
\and
{\it College of Engineering,
Chubu University,
1200 Matsumoto-cho, Kasugai, 487-8501, Japan}\\
\email{ozawa@isc.chubu.ac.jp}
}
\maketitle            
\begin{abstract}
In 1981, Takeuti introduced set theory based on quantum logic by constructing 
a model analogous to Boolean-valued models for Boolean logic.
He defined the quantum logical truth value for every sentence of set theory.
He showed that equality axioms do not hold, while axioms of ZFC set theory 
hold if appropriately modified with the notion of commutators.
Here, we consider the problem in Takeuti's quantum set theory that
De Morgan's laws do not hold for bounded quantifiers.
We construct a counter-example to De Morgan's laws for bounded quantifiers
in Takeuti's quantum set theory.
We redefine the truth value for the membership relation and 
bounded existential quantification to ensure that De Morgan's laws hold.
Then, we show that the truth value of every theorem of ZFC set theory 
is lower bounded by the commutator of constants therein as quantum transfer principle.

\end{abstract}
\setcounter{page}{1}
\section{Introduction}
Since quantum logic is an intrinsic logic governing observational propositions of 
quantum mechanics,  it is an intriguing problem to develop mathematics based on 
quantum logic.
In 1981, Takeuti \cite{Ta81} introduced quantum set theory for this purpose.
As a start, he constructed a model of set theory based on quantum logic
represented by the complete orthomodular lattice of projections on a
Hilbert space, which is isomorphic to the lattice of closed subspaces in 
the Hilbert space.  He defined the truth values for all sentences of set theory 
on the model assuming the Sasaki arrow for implication.
In order to make quantum counter part of ZFC axioms, he introduced the notion of
commutator in quantum logic, and  he showed that the  axioms of
ZFC hold in quantum set theory if appropriately modified by commutators of 
elements of the model, while equality axioms do not generally hold
in quantum set theory.
He showed that the real numbers in the model correspond to the 
observables of the system to be described. 

Following Takeuti's work, we explored the question how theorems of 
ZFC hold in quantum set theory \cite{07TPQ}.
We showed that every theorem of ZFC holds in quantum set theory 
with truth value greater than or equal to the commutator of elements of the model
appearing therein.  This result was extended to general complete orthomodular
lattices and to a general class of operations for implication in Ref.~\cite{17A2}.
Quantum set theory was effectively applied to quantum mechanics to extend 
the probabilistic interpretation from observational propositions to relations 
between observables \cite{16A2,17A1}.

In this paper, we consider the problem in Takeuti's quantum set theory that
De Morgan's laws do not hold for bounded quantifiers.
Let $\cH$ be a Hilbert space.
The quantum logic 
$\cQ$ on $\cH$ is represented by
the lattice of projections on $\cH$, which is a complete orthomodular lattice,
called the quantum logic on $\cH$.
The classical definition of implication, 
$P\Then Q=P^\perp\Or Q$, does not work since
the relation $P\Then Q=1$ and the order relation $P\le Q$ are not equivalent,
so that  the implication in quantum logic is, according to the majority view 
\cite{Urq83}, 
defined as the Sasaki arrow $\Then$,  a binary operation of $\cQ$ 
defined by $P\Then Q=P^\perp\Or (P\And Q)$.

Takeuti \cite{Ta81}, applying the method of Boolean-valued models to 
quantum logic $\cQ$, constructed the model $\VQ$ of quantum set theory.
He defined the $\cQ$-valued truth value $\val{\ph}$ of a sentence $\ph$
in the language of set theory.

In particular,  the truth values of bounded quantifications are directly defined as follows. 

(1) $ \val{(\forall x\in u)\, {\ph}(x)} 
= \Inf_{u'\in \dom(u)}
(u(u') \Then \val{\ph(u')})$.

(2) $ \val{(\exists x\in u)\, {\ph}(x)} 
= \Sup_{u'\in \dom(u)}
(u(u') \And \val{\ph(u')})$.

\noindent 
 Takeuti noted ``In Boolean valued universes, 
 $\val{(\forall x\in u)\ph(x)}=\val{\forall x(x\in u\Then \ph(x)}$
and 
$\val{(\exists x\in u)\ph(x)}=\val{\exists x(x\in u \And \ph(x)}$. 
But this is not the case for $\VQ$'' \cite[p.~315]{Ta81}.
However, it is problematic that he avoids the classical definition of
implication $P\Then Q=P\p\Or Q$ in the bounded universal 
quantification, 
whereas he still uses the classical definition of conjunction 
in the bounded existential quantification. 
Since 
the relation $P\And Q=(P\Then Q^{\perp})^{\perp}$ 
does not hold for the conjunction
$\And$ and the Sasaki arrow $\Then$, so that 
De Morgan's laws,

(3) $ \val{\neg(\forall x\in u)\, {\ph}(x)}=
 \val{(\exists x\in u)\,\neg {\ph}(x)}$,

(4) $ \val{\Not(\exists x\in u)\, {\ph}(x)}
=  \val{(\forall x\in u)\, \neg{\ph}(x)}$,

\noindent 
do not hold.  In fact, 
if $\cQ$ is not a Boolean algebra,  
we can construct a predicate $\ph(x)$ such that  
$\val{(\exists x\in u)\,\neg {\ph}(x)}=0$ but 
$\val{\neg(\forall x\in u)\, {\ph}(x)}>0$.

In this paper, we introduce a new binary operation $*$ by
$P*Q=(P\Then Q^{\perp})^{\perp}$ and redefine the truth values of 
membership relation and
bounded existential quantification as follows.

(5)  $ \val{u \in v} 
= \sup_{v' \in \cD(v)} (v(v') * \val{v'=u})$.

(6) $ \val{(\exists x\in u)\, {\ph}(x)} 
= \Sup_{u'\in \dom(u)}
(u(u') * \val{\ph(u')})$.

\noindent
Then, De Morgan's laws hold for bounded universal
quantification and bounded existential quantification.
Thus,  for the language of quantum set theory we can assume 
only negation,
conjunction, and bounded and unbounded universal quantification as
primitive, while disjunction, bounded and unbounded existential 
quantification are considered to be introduced by definition.

The operator $\ast$ was found by Sasaki \cite{Sas54},
and has been studied  as the Sasaki projection in connection 
with residuation theory, 
whereas the operation $\ast$ has not been used for defining
bounded quantifiers in quantum logic.  Its intuitive meaning and
significance will be discussed elsewhere.

We consider the commutator $\cm(u_1,\ldots,u_n)\in\cQ$ of elements 
$u_1,\ldots,u_n$ of $\VQ$ 
in order to explore how theorems of ZFC hold in the new interpretation 
for the model $\VQ$.
Then the following quantum transfer principle holds:
If a $\De_0$-formula $\ph(x_1,\ldots,x_n)$ of the language $\LL({\in})$ 
of set theory 
is provable in ZFC,  for every $u_1,\ldots,u_n\in\VQ$ we have
\[
\val{\phi(u_1,\ldots,u_n)}\ge \cm(u_1,\ldots,u_n).
\]

This paper is organized as follows.
Section 2 provides preliminaries on quantum logic, commutators, and
conditionals.
Section 3 introduces the model $\VQ$ and its interpretation that satisfies
De Morgan's laws.  We also discuss Takeuti's interpretation and construct
the above mentioned counterexample.  
Section 4 derives the quantum transfer principle for the new
interpretation.

\section{Preliminaries}
\subsection{Quantum logic}

Let $\cH$ be a Hilbert space.
For any subset $S\subseteq\cH$,
we denote by $S^{\perp}$ the orthogonal complement
of $S$.
Then, $S^{\perp\perp}$ is the closed linear span of $S$.
Let $\cC(\cH)$ be the set of all closed linear subspaces in
$\cH$. 
With the set inclusion ordering, 
the set $\cC(\cH)$ is a complete
lattice. 
The operation $M\mapsto M^\perp$ 
is  an orthocomplementation
on the lattice $\cC(\cH)$, with which $\cC(\cH)$ is a 
complete orthomodular lattices \cite[p.~65]{Kal83},
\ie, the orthocomplementation satisfies 

(C1)  if $P \le Q$ then $Q^{\perp}\le P^{\perp}$,

(C2) $P^{\perp\perp}=P$,

(C3) $P\Or P^{\perp}=1$ and $P\And P^{\perp}=0$,

\noindent where $0=\Inf\cQ$ and $1=\Sup\cQ$,
and the {\em orthomodular law}

(OM) if $P\le Q$ then $P\Or(P^{\perp}\And Q)=Q$.

We refer the reader to Kalmbach \cite{Kal83} for a standard textbook on
orthomodular lattices.

Denote by $\cB(\cH)$ the algebra of bounded linear
operators on $\cH$ and $\cQ(\cH)$ the set of projections on $\cH$.
We define the {\em operator ordering} on $\cB(\cH)$ by
$A\le B$ iff $(\ps,A\ps)\le (\ps,B\ps)$ for
all $\ps\in\cH$. 
For any $A\in\cB(\cH)$, denote by $\cR (A)\in\cC(\cH)$
the closure of the range of $A$, {\em i.e.,} 
$\cR(A)=(A\cH)^{\perp\perp}$.
For any $M\in\cC(\cH)$,
denote by $\cP (M)\in\cQ(\cH)$ the projection
of $\cH$ onto $M$.
Then, $\cR\cP (M)=M$ for all $M\in\cC(\cH)$
and $\cP\cR (P)=P$ for all $P\in\cQ(\cH)$,
and we have $P\le Q$ if and only if $\cR (P)\subseteq\cR (Q)$
for all $P,Q\in\cQ(\cH)$,
so that $\cQ(\cH)$ with the operator ordering is also a complete orthomodular
lattice isomorphic to $\cC(\cH)$. 
We consider $\cQ(\cH)$ as the {\em standard quantum logic} of $\cH$,
or the logic of observational propositions in quantum mechanics for
the system described by $\cH$ \cite{BvN36,Hus37}.
The lattice operations are characterized by 
$P\And Q={\mb{weak-lim}}_{n\to\infty}(PQ)^{n}$, 
$P^\perp=1-P$ for all $P,Q\in\cQ(\cH)$.

A non-empty subset of $\cQ(\cH)$ is called a {\em subalgebra} iff
it is closed under $\And $, $\Or$, and $\perp$.
A subalgebra $\cA$ of $\cQ(\cH)$ is said to be {\em complete} iff it has
the supremum and the infimum in $\cQ(\cH)$ of an arbitrary subset of $\cA$.

Let $\cA\subseteq\cB(\cH)$.
We denote by $\cA'$ the {\em commutant of 
$\cA$ in $\cB(\cH)$}, \ie, 
\[
\cA'=\{A\in\BH\mid \mb{$AB=BA$ for any $B\in\cA$}\}.
\]
A self-adjoint subalgebra $\cM$ of $\cB(\cH)$ is called a
{\em von Neumann algebra} on $\cH$ iff 
$\cM''=\cM$.
For any self-adjoint subset $\cA\subseteq\cB(\cH)$,
$\cA''$ is the von Neumann algebra generated by $\cA$.
We denote by $\cP(\cM)$ the set of projections in
a von Neumann algebra $\cM$.

We say that $P$ and $Q$ in $\cQ(\cH)$ {\em commute}, in  symbols
$P\commutes Q$, iff  $P=(P\And Q)\Or(P\And
Q^{\perp})$.  
For any $P,Q\in\cQ(\cH)$, we have 
$P\commutes Q$ iff $[P,Q]=0$, where $[P,Q]=PQ-QP$.

For any subset $\cA\subseteq\cQ(\cH)$,
the {\em commutant}  $\cA^{!}$ of $\cA$ in $\cQ(\cH)$  \cite[p.~23]{Kal83}
is defined by
\beq
\cA^{!}=
\{P\in\cQ(\cH)\mid P\commutes Q \mbox{ for all }
Q\in\cA\}.
\eeq
Then, $\cA^{!}$ is a complete subalgebra of $\cQ(\cH)$.
A {\em sublogic} of $\cQ(\cH)$ is a subset $\cA$ of
$\cQ(\cH)$ satisfying $\cA=\cA^{!!}$.
Any sublogic of $\cQ(\cH)$ will be called a {\em logic} on $\cH$. 
For any subset $\cA$ of a logic $\cQ$, the smallest 
logic including $\cA$ is 
$\cA^{!!}$ called the  {\em logic generated by
$\cA$}.
Then, a subset $\cQ \subseteq\cQ(\cH)$ is a logic on $\cH$ if
and only if $\cQ=\cP(\cM)$ for some von Neumann algebra
$\cM$ on $\cH$ \cite[Proposition 2.1]{07TPQ}.
A logic $\cQ$ on $\cH$ is a Boolean
algebra if and only if $P\commutes Q$  for all $P,Q\in\cQ$ \cite[pp.~24--25]{Kal83}

The center of a logic $\cQ$, denoted by $Z(\cQ)$, is the set of elements 
of $\cQ$ commute with every element of $\cQ$, \ie , $Z(\cQ)=\cQ^{!}\cap\cQ$. 
Then, it is easy to see that a subset 
 $\cA$ is a Boolean sublogic, or equivalently 
 a distributive sublogic, if and only if 
$\cA=\cA^{!!}\subseteq\cA^{!}$.
The center of $\cA^{!!}$ is given by $Z(\cA^{!!})=\cA^{!}\cap\cA^{!!}$.

\subsection{Commutators}
\label{se:CIQL}
Marsden \cite{Mar70} introduced the commutator $\com(P,Q)$ 
of two elements $P$ and $Q$ of an orthomodular lattice $\cQ$ by 
\beql{comM}
\com(P,Q)=(P\And Q)\Or(P\And Q\p)\Or(P\p\And Q)\Or(P\p\And Q\p).
\eeq
Bruns and Kalmbach \cite{BK73} generalized this notion 
to finite subsets of $\cQ$ by 
\beql{eq:comBK}
\com(\cF)=\Sup_{\theta:\cF\to\{\id,\perp\}}\Inf_{P\in\cF}P^{\theta(P)}
\eeq
for any finite subsets $\cF$ of $\cQ$, where
$\{\id,\perp\}$ stands for the set consisting of the identity operation $\id$ and the 
orthocomplementation~$\perp$.
Generalizing this notion to arbitrary subsets $\cA$ of $\cQ(\cH)$, Takeuti \cite{Ta81} defined
$\com(\cA)$ by
\beql{comT}
\com(\cA)=\Sup \{E\in\cA^{!} \mid P_{1}\And E\commutes P_{2}\And E
\mb{ for all }P_{1},P_{2}\in\cA\},
\eeq
for any subset $\cA$ of 
$\cQ(\cH)$.  
By Takeuti's definition it is not clear whether the commutator $\com(\cA)$ is determined
inside the logic $\cA^{!!}$ generated by $\cA$ or not, unlike the definition of $\com(\cF)$ 
for finite subsets $\cF$.  To resolve this problem, it was shown in 
Ref.~\cite[Theorem 2.5]{16A2} that 
$\com(\cA)\in \cA^{!}\cap \cA^{!!}$, and we obtain the relation
\beql{comO}
\com(\cA)=\Sup \{E\in\cA^{!}\cap \cA^{!!} \mid P_{1}\And E\commutes P_{2}\And E
\mb{ for all }P_{1},P_{2}\in\cA\},
\eeq
as an alternative definition for $\com(\cA)$.

We have the following characterizations of commutators 
\cite[Theorems 2.5, 2.6, Proposition 2.2]{07TPQ}:
For any $\cA\subseteq\cQ(\cH)$,
we have the following relations.
\begin{align}
\com(\cA)&=\cP\{\ps\in\cH\mid 
[P_1,P_2]P_3\psi=0\mb{ for all } P_1,P_2,P_3\in \cA\}.\\
\com(\cA)&=\cP\{\ps\in\cH\mid [A,B]\ps=0 \mb{ for all }A,B\in\cA''\}.
\end{align}
We refer the reader to Pulmannov\'{a} \cite{Pul85} and 
Chevalier \cite{Che89} for further results about commutators
in orthomodular lattices. 

\subsection{Conditionals}
\label{se:GIIQL}

In classical logic, the conditional operation $\Then$ is defined by negation $\perp$ and
disjunction $\Or$ as $P\Then Q=P^{\perp}\Or Q$.
In quantum logic there is well-known arbitrariness in choosing a binary operation for conditional.
Following Hardegree \cite{Har81}, we define a {\em quantum material conditional} 
on a logic $\cQ$ as a binary operation $\Then$ on $\cQ$ definable 
by an ortholattice polynomial $p(x,y)$  as $P\Then Q=p(P,Q)$ for all $P,Q\in\cQ$ satisfying the following 
``minimum implicative conditions'':
\benum
\item[(E)]  $P\Then Q=1$ if and only if $P\le Q$. 
\item[(MP)] ({\it modus ponens}) $P\And(P\Then Q)\le Q$.
\item[(MT)] ({\it modus tollens}) $Q^{\perp}\And (P\Then Q) \le P^{\perp}$.
\eenum
  
  Hardegree \cite{Har81} showed that there are exactly three polynomially definable 
  material conditionals:
\benum
\item[(S)] (Sasaki conditional) $P\Then{}_{S}Q: =P^{\perp}\Or(P\And Q)$, 
\item[(C)] (Contrapositive Sasaki conditional) $P\Then{}_{C}Q: =(P\Or Q)^{\perp}\Or Q$,
\item[(R)] (Relevance conditional) $P\Then{}_{R}Q: =(P\And Q)\Or(P^{\perp}\And Q)\Or(P^{\perp}\And Q^{\perp})$.
\eenum

Following Takeuti \cite{Ta81} we adopt the Sasaki arrow,
the most favorable according to the majority view \cite{Urq83},
 as the conditional $\Then$ for a
logic $\cQ$, \ie, $P\Then Q=P\p\Or(P\And Q)$.
The logical equivalence $\Iff$ is defined
by
\beq
P\Iff Q = (P\Then Q)\And(Q\Then P).
\eeq

In Boolean logic, implication and conjunction are associated by
the relation $P\And Q=(P\Then Q^{\perp})\p$, and this relation plays
an essential role in the duality between bounded universal quantification
$(\forall x\in A)\ph(x)$ and bounded existential quantification
$(\exists x\in A)\ph(x)$.  
In order to keep the above duality in quantum set theory, 
we  introduce the binary  operation $*$ dual to $\Thenj$ by
\beq
P*Q=(P\Thenj Q\p)\p.
\eeq
We have the following relations
\begin{align}
P\Then Q&=(P\And Q)\Or(P^{\perp}\And Q)\Or(P^{\perp}\And Q^{\perp})\Or(P\p\And\com(P,Q)\p).\\
P *Q&=(P\And Q)\Or (P\And \com(P,Q)\p).
\label{eq:star}
\end{align}

The following proposition is useful in later discussions \cite[Proposition 2.4]{07TPQ},
\cite[Proposition 3.1]{17A2}.

\begin{Proposition}\label{th:logic}
Let $\cQ$ be a  logic on $\cH$.
The following hold.

(i)  If $P_{\al}\in\cQ$ and 
$P_{\al}\commutes Q$ for all $\al$, then
$(\Sup_{\al}P_{\al})\commutes Q$, 
$(\Inf_{\al}P_{\al})\commutes Q$, 
and
$Q \And (\Sup_{\al}P_{\al})=\Sup_{\al}(Q\And
P_{\al}).$

(ii) If $P_1,P_2\commutes Q$, then 
$(P_1\Then P_2)\And Q
=[(P_1\And Q)\Then(P_2\And Q)]\And Q$.

(iii) If $P_1,P_2\commutes Q$, then 
$(P_1* P_2)\And Q
=[(P_1\And Q)*(P_2\And Q)]\And Q$.
\end{Proposition}

\section{Quantum set theory}
\label{se:UQ}

We denote by $\V$ the universe of the Zermelo-Fraenkel set theory
with the axiom of choice (ZFC).
Let $\cL(\in)$ be the language of first-order theory with equality 
augmented by a connective $\Implies$,
a binary relation symbol $\in$, bounded quantifier symbols $\forall x\in y$, $\exists x \in y$, 
and no constant symbols.
For any class $U$, the language $\cL(\in,U)$ is the one obtained by adding a name 
for each element of $U$.
We take the symbols $\Not$, $\And$, $\Then$, $\forall x\in y$, and $\forall x$
as primitive, and the symbols $\Or$, 
$\exists x\in y$, and $\exists x$ as derived
symbols by defining:
\benum
\item $\ph\Or\ps=\Not(\Not \ph\And \Not\ps)$,
\item $\exists x\in y\,\ph(x)=\Not(\forall x\in y\,\Not\ph(x)),$
\item $\exists x\ph(x)=\Not(\forall x\,\Not\ph(x)).$
\eenum

To each statement $\ph$ of $\LL(\in,U)$, the satisfaction
relation
$\bracket{U,\in} \models \ph$ is defined by the following recursive rules:
\begin{enumerate}
\item $ \bracket{U,\in} \models u\in v
\iff u\in v.$
\item $ \bracket{U,\in} \models u = v
\iff u = v.$ 
\item $ \bracket{U,\in} \models \Not \ph \iff  \bracket{U,\in}
\models
\ph 
\mbox{ does not hold}$. 
\item $ \bracket{U,\in} \models \ph_1 \And \ph_2 
\iff \bracket{U,\in}
\models \ph_1 
\mbox{ and } \bracket{U,\in} \models \ph_2$.
\item $ \bracket{U,\in} \models \ph_1 \Then \ph_2 
\iff \mb{ if }\bracket{U,\in}\models \ph_1 
\mbox{ then } \bracket{U,\in} \models \ph_2$.
\item $ \bracket{U,\in} \models  (\forall x\in u)\,\ph(x) \iff
\bracket{U,\in} \models \ph(u') \mbox{ for all } u' \in u$.  
\item $ \bracket{U,\in} \models  (\forall x)\,\ph(x) \iff
\bracket{U,\in} \models \ph(u) \mbox{ for all } u \in U$.
\end{enumerate}
Our assumption that $\V$ satisfies ZFC
means that if $\ph(x_1,\ldots,x_n)$ is provable in 
ZFC,  \ie, 
$\mb{ZFC}\vdash \ph(x_1,\ldots,x_n)$, then 
$\bracket{\V,\in}\models \ph(u_1,\ldots,u_n)$ for 
any formula $\ph(x_1,\ldots,x_n)$ of $\LL(\in)$ and
all $u_1,\ldots,u_n\in \V$. 

In what follows let $\cQ$ be a logic on $\cH$. 
For each ordinal $ {\al}$, let
\beq
\V_{\al}^{(\cQ)} = \{u|\ u:\dom(u)\to \cQ \mbox{ and }
(\exists \be<\al)
\dom(u) \subseteq V_{\be}^{(\cQ)}\}.
\eeq
The {\em $\cQ$-valued universe} $\VL$ is defined
by 
\beq
  \VL= \bigcup _{{\al}{\in}\mbox{On}} V_{{\al}}^{(\cQ)},
\eeq
where $\mbox{On}$ is the class of all ordinals. 
For every $u\in\VQ$, the rank of $u$, denoted by
$\rank(u)$,  is defined as the least $\al$ such that
$u\in \VQ_{\al+1}$.
It is easy to see that if $u\in\dom(v)$ then 
$\rank(u)<\rank(v)$.

An induction on rank argument leads to the following \cite{Bel05}.
\bTheorem[Induction Principle for $\VQ$]
For any predicate $\ph(x)$,
\[
\forall u\in \VQ[\forall u'\in\dom(u)\ph(u')\Then\ph(u)]\Then \forall u\in\VQ\ph(u)
\]
\eTheorem

For any $u,v\in\VL$, the $\cQ$-valued truth values 
$\vval{u=v}$ and $\vval{u\in v}$ of
atomic formulas $u=v$ and $u\in v$ are assigned
by the following rules recursive in rank.
\begin{enumerate}[(i)]\itemsep=0in
\setcounter{enumi}{3}
\item $\vval{u = v}
= \inf_{u' \in  \cD(u)}(u(u') \Thenj
\vval{u'  \in v})
\And \inf_{v' \in   \cD(v)}(v(v') 
\Thenj \vval{v'  \in u})$.
\item $ \vval{u \in v} 
= \sup_{v' \in \cD(v)} (v(v')* \vval{u =v'})$.
\end{enumerate}

To each statement $\ph$ of $\cL(\in,\VL)$ 
we assign the $\cQ$-valued truth value $ \vval{\ph}$ by the following
rules.
\begin{enumerate}[(i)]\itemsep=0in
\setcounter{enumi}{5}
\item $ \vval{\Not\ph} = \vval{\ph}^{\perp}$.
\item $ \vval{\ph_1\And\ph_2} 
= \vval{\ph_{1}} \And \vval{\ph_{2}}$.
\item $ \vval{\ph_1\rightarrow\ph_2} 
= \vval{\ph_{1}} \Thenj \vval{\ph_{2}}$.
\item $ \vval{(\forall x\in u)\, {\ph}(x)} 
= \Inf_{u'\in \dom(u)}
(u(u') \Thenj \vval{\ph(u')})$.
\item $ \vval{(\forall x)\, {\ph}(x)} 
= \Inf_{u\in \VL}\vval{\ph(u)}$.
\end{enumerate}

By the definitions of derived logical symbols, (i)--(iii),  we have the following relations.
\begin{enumerate}[(i)]\itemsep=0in
\setcounter{enumi}{10}
\item $ \vval{\ph_1\Or\ph_2} 
= \vval{\ph_{1}} \Or \vval{\ph_{2}}$.
\item $ \vval{(\exists x\in u)\, {\ph}(x)} 
= \Sup_{u'\in \dom(u)}(u(u') * \vval{\ph(u')})$.
\item $ \vval{(\exists x)\, {\ph}(x)} 
= \Sup_{u\in \VL}\vval{\ph(u)}$.
\end{enumerate}

To see (xii),   we have
\begin{align*}
\vval{(\exists x\in u)\, {\ph}(x)} 
&=\vval{\Not(\forall x\in u\,\Not\ph(x))}\\
&=(\Inf_{u'\in \dom(u)}
u(u') \Thenj \vval{\Not\ph(u')})\p\\
&=(\Inf_{u'\in \dom(u)}
u(u') \Thenj \vval{\ph(u')}\p)\p\\
&=\Sup_{u'\in \dom(u)}
(u(u') \Thenj \vval{\ph(u')}\p)\p\\
&=\Sup_{u'\in \dom(u)}
(u(u') * \vval{\ph(u')}).
\end{align*}

Note that according to the above, we have the following relations
\begin{enumerate}[(i)]\itemsep=0in
\setcounter{enumi}{13}
\item $\vval{u = v}=\vval{\forall x\in u(x\in v)\And \forall x\in v(x\in u)}$,
\item $\vval{u  \in v}=\vval{\exists x\in v(x=u)}.$
\end{enumerate}

We also have the following relations satisfying De Morgan's laws:
\begin{enumerate}[(i)]\itemsep=0in
\setcounter{enumi}{15}
\item$\vval{\Not(\ph_1\And\ph_2)} =\vval{\Not \ph_1\Or \Not \ph_2},$
\item $\vval{\Not(\ph_1\Or \ph_2)}=\vval{\Not \ph_1\And \Not \ph_2},$
\item $\vval{\Not(\forall x\in u\,\ph(x))}=\vval{\exists x\in u\,(\Not \ph(x))},$
\item $\vval{\Not(\exists x\in u\,\ph(x))}=\vval{\forall x\in u\,(\Not \ph(x))},$
\item $\vval{\Not(\forall x\,\ph(x))}=\vval{\exists x\,(\Not \ph(x))},$
\item $\vval{\Not(\exists x\,\ph(x))}=\vval{\forall x\,(\Not \ph(x))}.$
 \end{enumerate}

A formula in $\cL(\in)$ is called a {\em
$\De_{0}$-formula}  iff it has no unbounded quantifiers
$\forall x$ nor $\exists x$.
The following theorem holds.

\sloppy
\bTheorem[$\De_{0}$-Absoluteness Principle]
\label{th:Absoluteness}
\sloppy  
For any $\De_{0}$-formula 
${\ph} (x_{1},{\ldots}, x_{n}) $ 
of $\cL(\in)$ and $u_{1},{\ldots}, u_{n}\in \VQ$, 
we have
\[
\vval{\ph(u_{1},\ldots,u_{n})}=
\val{\ph(u_{1},\ldots,u_{n})}_{\cQ(\cH)}.
\]
\eTheorem
\bProof 
The assertion is proved by the induction on the complexity
of formulas and the rank of elements of $\VQ$.
Let $u, v\in\VL$.
By induction hypothesis, for any $u'\in\dom(u)$ and $v'\in\dom(v)$
we have
$\vval{u'\in w}=\vvall{u'\in w}$,
$\vval{v'\in w}=\vvall{v'\in w}$,
and $\vval{w=v'}=\vvall{w=v'}$
for all $w\in \VQ$.
Thus, 
\beqas
\vval{u=v}
&=&\Inf_{u'\in\dom(u)}(u(u')\Thenj\vval{u'\in v})
\And
\Inf_{v'\in\dom(v)}(v(v')\Thenj\vval{v'\in u})\\
&=&\Inf_{u'\in\dom(u)}(u(u')\Thenj\vvall{u'\in v})
\And
\Inf_{v'\in\dom(v)}(v(v')\Thenj\vvall{v'\in u})\\
&=&
\vvall{u=v},
\eeqas
and  we also have
\beqas
\vval{u\in v}
&=&
\Sup_{v'\in\dom(v)}(v(v')*\vval{u=v'})\\  
&=&
\Sup_{v'\in\dom(v)}(v(v')*\vvall{u=v'})\\ 
&=&
\vvall{u\in v}.
\eeqas
Thus, the assertion holds for atomic formulas.
Any induction step adding a logical symbol works
easily, even when bounded quantifiers are concerned,
since the ranges of the supremum and the infimum 
are common for evaluating $\vval{\cdots}$ and 
$\vvall{\cdots}$. 
\eProof

Henceforth, 
for any $\De_{0}$-formula 
${\ph} (x_{1},{\ldots}, x_{n}) $
and $u_1,\ldots,u_n\in\VQ$,
we abbreviate $\val{\ph(u_{1},\ldots,u_{n})}=
\vval{\ph(u_{1},\ldots,u_{n})}$,
which is the common $\cQ(\cH)$-valued truth value for 
$u_{1},\ldots,u_{n}\in\VQ$.

The universe $\V$  can be embedded in
$\VQ$ by the following operation 
$\vee:v\mapsto\check{v}$ 
defined by the $\in$-recursion: 
for each $v\in\V$, $\check{v} = \{\check{u}|\ u\in v\} 
\times \{1\}$. 
Then we have the following.
\bTheorem[$\De_0$-Elementary Equivalence Principle]
\label{th:2.3.2}
\sloppy
Let ${\ph} (x_{1},{\ldots}, x_{n}) $ be a 
$\De_{0}$-for\-mu\-la  of $\cL(\in)$.
For any $u_{1},{\ldots}, u_{n}\in V$,
we have
$$
\bracket{\V,\in}\models {\ph}(u_{1},{\ldots},u_{n})
\quad\mbox{if and only if}\quad
\val{\ph(\check{u}_{1},\ldots,\check{u}_{n})}=1.
$$
\eTheorem
\bProof
Let ${\bf 2}$ be the sublogic such that ${\bf 2}=\{0,1\}$.
Then, by induction it is easy to see that 
$
\bracket{\V,\in}\models  {\ph}(u_{1},{\ldots},u_{n})
\mbox{ if and only if }
\val{\ph(\check{u}_{1},\ldots,\check{u}_{n})}_{\bf 2}=1
$
for any ${\ph} (x_{1},{\ldots}, x_{n})$ in $\LL(\in)$, 
and this is
equivalent  to $\valj{\ph(\check{u}_{1},\ldots,\check{u}_{n})}=1$
for any $\De_{0}$-formula ${\ph} (x_{1},{\ldots}, x_{n})$ 
by the $\De_0$-absoluteness principle.
\eProof

Instead of (v) and (xii),
Takeuti \cite{Ta81} defined the truth values of membership relation and
existential quantification as follows.
\benum 
\item[(v')] $ \val{u \in v} 
= \Sup_{v' \in \cD(v)} (v(v')\And \val{u =v'})$.
\item[(xii')] $ \val{(\exists x\in u)\, {\ph}(x)} 
= \Sup_{u'\in \dom(u)}(u(u') \And \val{\ph(u')})$.
\eenum
  In this case, De Morgan's laws do not hold in general as follows.
  
Suppose that $\cQ$ is not a Boolean algebra.
Then, there exists a pair
$P_0,Q_0\in\cQ$ such that 
$P_0$ does not commute with $Q_0$, so that $\com(P_0,Q_0)\p>0$.
Let $E=\com(P_0,Q_0)\p$, $P=P_0\And E$, and $Q=Q_0\And E$.  
If $P=0$ then $P_0=P_0\And \com(P_0,Q_0)$ so that
$P_0\commutes Q_0$, a contradiction.  Thus, $P\ne 0$.  
We also have that $P\And Q=P_0\And Q_0 \And \com(P_0,Q_0)\p=0$,
so that $P\And Q=0$.
Let $u=\{\av{\ck{0},P}\}$ and $v=\{\av{\ck{0},Q}\}$. 
Consider the formula $\ph(x)=\Not(x\in v)$.  Then, we can show 
\beql{eq:non-DM}
\val{\Not (\forall x\in u)\ph(x)}>\val{(\exists x\in u)\Not \ph(x)}=0.
\eeq 
In fact, we have
\begin{align*}
\val{(\exists x\in u)\Not \ph(x)}
&= \Sup_{u'\in \dom(u)}(u(u') \And \val{\Not\ph(u')})\\
&= u(\ck{0}) \And \val{\Not\Not(\ck{0}\in v)}\\
&= u(\ck{0}) \And \val{\ck{0}\in v}\\
&=u(\ck{0}) \And \Sup_{v' \in \dom(v)} (v(v')\And \val{\ck{0} =v'})\\
&=u(\ck{0}) \And (v(\ck{0})\And \val{\ck{0} =\ck{0}})\\
&=u(\ck{0}) \And v(\ck{0})\\
&=P\And Q\\
&=0.
\end{align*}
\noindent
Similarly we have
\begin{align*}
\val{\Not (\forall x\in u)\ph(x)}
&=\val{(\forall x\in u)\ph(x)}\p\\
&=(\Inf_{u'\in \dom(u)}(u(u')\Then\val{\ph(u')}))\p\\
&=(u(\ck{0})\Then\val{\ph(\ck{0})})\p\\
&=u(\ck{0})*\val{\ph(\ck{0})}\p\\
&= u(\ck{0}) * \val{\Not(\ck{0}\in v)}\p\\
&= u(\ck{0}) * \val{\ck{0}\in v}\\
&=u(\ck{0}) * \Sup_{v' \in \dom(v)} (v(v')\And \val{\ck{0} =v'})\\ 
&=u(\ck{0}) * (v(\ck{0})\And \val{\ck{0} =\ck{0}})\\ 
&=u(\ck{0}) * v(\ck{0})\\
&=P * Q\\
&=(P\And Q)\Or(P\And\com(P,Q)\p)\\
&=P.
\end{align*}
Since $P\not=0$, \Eq{eq:non-DM} follows.

Thus, if $\cQ$ is not a Boolean algebra, 
 there exists a predicate  
$\ph(x)$ such that  $\val{(\exists x\in u)\,\neg {\ph}(x)}=0$ but 
$\val{\neg(\forall x\in u)\, {\ph}(x)}>0$.

\section{Transfer principle}
\label{se:ZFC}\label{se:TPQ}

In this section, we investigate the transfer principle that gives any $\De_0$-formula 
provable in ZFC a lower bound for its truth value, which is determined by the degree 
of the commutativity of the elements of $\VQ$ appearing in the formula as constants.
The results in this section was obtained in Ref.~\cite{07TPQ} for Takeuti's original
formulation.  Here, we extends the argument in a self-contained manner to the present 
formulation, in which De Morgan's laws hold for bounded quantifiers.

For $u\in\VQ$, we define the {\em support} 
of $u$, denoted by $L(u)$, by transfinite recursion on the 
rank of $u$ by the relation
\beq
L(u)=\bigcup_{x\in\dom(u)}L(x)\cup\{u(x)\mid x\in\dom(u)\} \cup \{0\}.
\eeq
For $\cA\subseteq\VQ$ we write 
$L(\cA)=\bigcup_{u\in\cA}L(u)$ and
for $u_1,\ldots,u_n\in\VQ$ we write 
$L(u_1,\ldots,u_n)=L(\{u_1,\ldots,u_n\})$.
Then, we obtain the following characterization of
subuniverses of $V^{(\cQ(\cH))}$.

\begin{Proposition}\label{th:sublogic}
Let $\cQ$ be a logic on $\cH$ and $\al$ an
ordinal. For any $u\in V^{(\cQ(\cH))}$, we have
$u\in\VL_{\al}$  if and only if
$u\in V^{(\cQ(\cH))}_{\al}$ and $L(u)\subseteq\cQ$.  
In particular, $u\in\VL$ if and only if
$u\in\VQH$ and $L(u)\subseteq\cQ$. 
Moreover, $\rank(u)$ is the least $\al$ such 
that $u\in \VQH_{\al+1}$ for  any $u\in\VL$.
\end{Proposition}
\bProof Immediate from transfinite induction on
$\al$.
\eProof

Let $\cA\subseteq\VQ$.  The {\em commutator
of $\cA$}, denoted by $\cm(\cA)$, is 
defined by  
\beq
\cuniv(\cA)=\com (L(\cA)).
\eeq
For any $u_1,\ldots,u_n\in\VQ$, we write
$\cuniv(u_1,\ldots,u_n)=\cuniv(\{u_1,\ldots,u_n\})$.

Let $u\in\VQ$ and $p\in\cQ$.
The {\em restriction} $u|_p$ of $u$ to $p$ is defined by
the following transfinite recursion:
\beqas
u|_p=\{\av{x|_p,u(x)\And p}\mid x\in\dom(u)\}\cup\{\av{u,0}\}.
\eeqas
The last term $\{\av{u,0}\}$ has no essential role but 
ensures the well-definedness
of the function $u|_p:\dom(u|_p)\to \cQ$.

\begin{Proposition}\label{th:L-restriction}
For any $\cA\subseteq \VQ$ and $p\in\cQ$, 
we have 
\beq
L(\{u|_p\mid u\in\cA\})=L(\cA)\And p.
\eeq
\end{Proposition}
\bProof
By induction, it is  easy to see the relation
$
L(u|_p)=L(u)\And p,
$
so that the assertion follows easily.
\eProof

Let $\cA\subseteq\VQ$.  The {\em logic
generated by $\cA$}, denoted by $\cQ(\cA)$, is  defined by 
\beq
\cQ(\cA)=L(\cA)^{!!}.
\eeq
For $u_1,\ldots,u_n\in\VQ$, we write
$\cQ(u_1,\ldots,u_n)=\cQ(\{u_1,\ldots,u_n\})$.

\begin{Proposition}\label{th:range}
For any $\De_0$-formula $\ph(x_1,\ldots,x_n)$ in
$\LL(\in)$ and $u_1,\cdots,u_n\in\VQ$,
we have $\val{\ph(u_1,\ldots,u_n)}\in\cQ(u_1,\ldots,u_n)$.
\end{Proposition}
\bProof
Let $\cA=\{u_1,\ldots,u_n\}$.
Since $L(\cA)\subseteq\cQ(\cA)$, it follows from
Proposition \ref{th:sublogic} that $u_1,\ldots,u_n\in
V^{(\cQ(\cA))}$.
By the $\De_0$-absoluteness
principle, we have 
$\val{\ph(u_1,\ldots,u_n)}=
\val{\ph(u_1,\ldots,u_n)}{}_{\cQ(\cA)}\in \cQ(\cA)$.
\eProof

\begin{Proposition}\label{th:commutativity}
For any 
$\De_{0}$-formula ${\ph} (x_{1},{\ldots}, x_{n})$ 
in $\LL(\in)$ and $u_{1},{\ldots}, u_{n}\in\VQ$, if 
$p\in L(u_1,\ldots,u_n)^{!}$, then 
$p\commutes \val{\ph(u_1,\ldots,u_n)}$
and $p\commutes \val{\ph(u_1|_p,\ldots,u_n|_p)}$.
\end{Proposition}
\bProof
Let $u_{1},{\ldots}, u_{n}\in\VQ$.
If $p\in L(u_1,\ldots,u_n)^{!}$, then
$p\in \cQ(u_1,\ldots,u_n)^{!}$.  From Proposition 
\ref{th:range},
$\val{\ph(u_1,\ldots,u_n)}\in\cQ(u_1,\ldots,u_n)$,
so that $p\commutes \val{\ph(u_1,\ldots,u_n)}$.
From Proposition \ref{th:L-restriction},
$L(u_1|_p,\ldots,u_n|_p)=L(u_1,\ldots,u_n)\And p$,
and hence $p\in L(u_1|_p,\ldots,u_n|_p)^{!}$, so that
$p\commutes \val{\ph(u_1|_p,\ldots,u_n|_p)}$.
\eProof

We define the binary relation $x_1\subseteq x_2$ by
$\forall x\in x_1(x\in x_2)$.
Then, by definition for  any $u,v\in\VQ$ we have
\beq
\val{u\subseteq v}=
\Inf_{u'\in\dom(u)}(u(u')\Thenj \val{u'\in v}),
\eeq
and we have $\val{u=v}=\val{u\subseteq v}
\And\val{v\subseteq u}$.

\begin{Proposition}\label{th:restriction-atom}
For any $u,v\in\VQ$ and $p\in L(u,v)^{!}$, we have
the following relations.

(i) $\val{u|_p\in v|_p}=\val{u\in v}\And p$.

(ii) $\val{u|_p\subseteq v|_p}\And p
=\val{u\subseteq v}\And p$.

(iii) $\val{u|_p= v|_p}\And p =\val{u= v}\And p$
\end{Proposition}
\bProof
We prove the relations by induction on the ranks of 
$u,v$.  If $\rank(u)=\rank(v)=0$, then $\dom(u)=\dom(v)
=\emptyset$, so that the relations trivially hold.
Let $u,v\in\VQ$ and $p\in L(u,v)^{!}$.
To prove (i),
let $v'\in\dom(v)$. 
Then, we have $p\commutes v(v')$ by the assumption on $p$.
By induction hypothesis, we have also 
$\val{u|_p=v'|_p}\And p=\val{u=v'}\And p$.
By Proposition \ref{th:commutativity}, we have 
$p\commutes \val{u=v'}$, so that
$v(v'), \val{u=v'}\in\{p\}^{!}$.
From \Eq{star} we have $(v(v')\And p)*\val{u|_p=v'|_p}\le p$,
and hence we have 
\begin{align*}
(v(v')\And p)*\val{u|_p=v'|_p}
&=
(v(v')\And p)*(\val{u|_p=v'|_p}\And p)\\
&=
(v(v')\And p)*(\val{u=v'}\And p)\\
&=
(v(v')*\val{u=v'})\And p.
\end{align*}
Thus,  we  have
\beqas
\val{u|_p\in v|_p}
&=&\Sup_{v'\in\dom(v|_p)}
(v|_p(v')*\val{u|_p=v'})\\
&=&
\Sup_{v'\in\dom(v)}
(v|_p(v'|_p)*\val{u|_p=v'|_p})\\
&=&
\Sup_{v'\in\dom(v)}
[(v(v')\And p)*\val{u|_p=v'|_p}]\\
&=&
\Sup_{v'\in\dom(v)}
[(v(v')*\val{u=v'})\And p]\\
&=&
\left(\Sup_{v'\in\dom(v)}
(v(v')* \val{u=v'})\right)\And p,
\eeqas  
where the last equality follows from Proposition \ref{th:logic} (i).
Thus, by definition of $\val{u\in v}$ we obtain
the relation $\val{u|_p\in v|_p}=\val{u\in v}\And p$,
and relation (i) has been proved.
To prove (ii), let $u'\in\dom(u)$.
Then, we have $\val{u'|_p\in v|_p}=\val{u'\in v}\And p$
by induction hypothesis.
Thus, we have
\beqas
\val{u|_p\subseteq v|_p}
&=&
\Inf_{u'\in\dom(u|_p)}(u|_p(u')\Thenj\val{u'\in v|_p})\\
&=&
\Inf_{u'\in\dom(u)}(u|_p(u'|_p)\Thenj\val{u'|_p\in v|_p})\\
&=&
\Inf_{u'\in\dom(u)}
[(u(u')\And p)\Thenj(\val{u'\in v}\And p)].
\eeqas
We  have $p\commutes u(u')$ by
assumption on $p$, and $p\commutes\val{u'\in v}$
by Proposition \ref{th:commutativity},
so that
$p\commutes u(u')\Thenj\val{u'\in v}$ and
$p\commutes (u(u')\And p)\Thenj(\val{u'\in v}\And p)$.
Thus, by Proposition \ref{th:logic} 
we have
\beqas
p\And\val{u|_p\subseteq v|_p}
&=&
p\And\Inf_{u'\in\dom(u)}
[(u(u')\And p)\Thenj(\val{u'\in v}\And p)]\\
&=&
p\And\Inf_{u'\in\dom(u)}(u(u')\Thenj\val{u'\in v})\\
&=&
p\And\val{u\subseteq v}.
\eeqas
Thus, we have proved  relation (ii).  
Relation (iii) follows easily from relation (ii).
\eProof

We have the following theorem.
\bTheorem[$\De_0$-Restriction Principle]
\label{th:V-restriction}
For any $\De_{0}$-formula ${\ph} (x_{1},{\ldots}, x_{n})$ in
$\LL(\in)$ and $u_{1},{\ldots}, u_{n}\in\VQ$, if 
$p\in L(u_1,\ldots,u_n)^{!}$, then 
$\val{\ph(u_1,\ldots,u_n)}\And p=
\val{\ph(u_1|_p,\ldots,u_n|_p)}\And p$.
\eTheorem
\bProof
We shall write $\vec{u}=(u_1,\ldots,u_n)$ and 
$\vec{u}|_p=(u_1|_p,\ldots,u_n|_p)$.
We prove the assertion by induction on 
the complexity of  ${\ph} (x_{1},{\ldots},x_{n})$.
From Proposition \ref{th:restriction-atom}, the assertion
holds for atomic formulas.
Thus, it suffices to consider the following induction steps: 
(i) $\ph \THEN \Not\ph$, (ii) $\ph_1,\ph_2 \THEN \ph_1\And\ph_2$
(iii) $\ph_1,\ph_2 \THEN \ph_1\Then\ph_2$, 
(iv) $\{\ph(x)\mid x\in\dom(u)\} \Then 
\Inf_{x\in\dom(u)}\ph(x)$.

(i) If $a\commutes p$, the relation
\beql{not}
a^{\perp}\And p=(a\And p)^{\perp}\And p
\eeq
follows easily.  Let $p\in L(\vec{u})^{!}$.
Suppose $\val{\ph(\vec{u})}\And p=\val{\ph(\vec{u}|_p)}\And p$.  
From \Eq{not} we have
\begin{align*}
\val{\ph(\vec{u})}\p\And p
&=(\val{\ph(\vec{u})}\And p)\p\And p\\
&=(\val{\ph(\vec{u}|_p)}\And p)\p\And p\\
&=\val{\ph(\vec{u}|_p)}\p\And p,
\end{align*}
so that we have
$$
\val{\Not\ph(\vec{u})}\And p=\val{\Not\ph(\vec{u}|_p)}\And p.
$$

(ii) Let $p\in L(\vec{u})^{!}$.
Suppose $\val{\ph_j(\vec{u})}\And p=\val{\ph_j(\vec{u}|_p)}\And p$
for $j=1,2$.
Then, it follows easily from associativity of $\And$, we have
\[
\val{\ph_1(\vec{u})\And\ph_2(\vec{u})}\And p
=\val{\ph_1(\vec{u}|_p)\And\ph_2(\vec{u}|_p)}\And p.
\]

(iii) Recall the relation
\[
(a\Thenj b)\And p=[(a\And p)\Thenj (b\And p)] 
\And p
\] 
for  all $a,b\in\{p\}^{!}$ as shown in Proposition \ref{th:logic} (ii).
Let $p\in L(\vec{u})^{!}$.
Suppose $\val{\ph_j(\vec{u})}\And p=\val{\ph_j(\vec{u}|_p)}\And p$
for $j=1,2$.
It follows from the above relation and the induction hypothesis that
\begin{align*}
\val{\ph_1(\vec{u})\Then\ph_2(\vec{u})}\And p
&=
[(\val{\ph_1(\vec{u})}\And p)\Then(\val{\ph_2(\vec{u})}\And p)]\And p\\
&=
[(\val{\ph_1(\vec{u}|_p)}\And p)\Then(\val{\ph_2(\vec{u}|_p)}\And p)]\And p\\
&=
(\val{\ph_1(\vec{u}|_p)}\Then\val{\ph_2(\vec{u}|_p)})\And p,
\end{align*}
so that we have
\[
\val{\ph_1(\vec{u})\Then\ph_2(\vec{u})}\And p=
\val{\ph_1(\vec{u}|_p)\Then\ph_2(\vec{u}|_p)}\And p.
\]

(iv)  Note that the relation
\[
(\Inf_{\al} P_{1,\al}\Then P_{2,\al})\And Q=
(\Inf_{\al} (P_{1,\al}\And Q)\Then(P_{2,\al}\And Q))\And Q
\] 
holds if $P_{j,\al}\commutes Q$ for $j=1,2$, 
which follows from  Proposition \ref{th:logic} (i) and (ii).
Suppose 
$\val{\ph_j(u)}\And p=\val{\ph_j(u|_p)}\And p$ for $j=1,2$
for any $u\in\VQ$ and $p\in L(u)^{!}$. 
Suppose $u\in\VQ$ and $p\in L(u)^{!}$.
Let $u'\in\dom(u)$.   Since $L(u')\subseteq L(u)$, we have $p\in L(u')^!$.
It follows that 
\[
\val{\ph_j(u')}\And p=\val{\ph_j(u'|_p)}\And p
\quad\mb{and}\quad p\commutes\val{\ph(u')}, 
\val{\ph(u'|_p)}
\]
for all $u'\in\dom(u)$.
Thus, we have

\begin{align*}
\val{(\forall x\in u)\ph(x)}\And p
&=
\left(\Inf_{u'\in\dom(u)}(u(u')\Then\val{\ph(u')})\right)\And p\\
&=
\Inf_{u'\in\dom(u)}[(u(u')\Then\val{\ph(u')})\And p]\\
&=
\Inf_{u'\in\dom(u)}\{[(u(u')\And p)\Then(\val{\ph(u')}\And p)]\And p\}\\
&=
\Inf_{u'\in\dom(u|_p)}\{[u|_p(u')\And p\Then(\val{\ph(u')}\And p)]\And p\}\\
&=
\Inf_{u'\in\dom(u|_p)}\{[u|_p(u')\Then(\val{\ph(u')})]\And p\}\\
&=
\left(\Inf_{u'\in\dom(u|_p)}(u|_p(u')\Then\val{\ph(u')})\right)\And p.
\end{align*}
It follows that 
$$
\val{(\forall x\in u)\ph(x)}\And p=
\val{(\forall x\in u|_p)\ph(x)}\And p.
$$
\eProof

Now,  we obtain the following transfer principle for bounded theorems of ZFC
in the new truth-value assignments for membership and existential quantifications
to fully satisfy De Morgan's laws.

\bTheorem[$\De_{0}$-ZFC  Transfer Principle]
For any $\De_{0}$-formula ${\ph} (x_{1},{\ldots}, x_{n})$ 
of $\cL(\in)$ and $u_{1},{\ldots}, u_{n}\in\VQ$, if 
${\ph} (x_{1},{\ldots}, x_{n})$ is provable in ZFC, then
we have
\beq
\val{\ph({u}_{1},\ldots,{u}_{n})}\ge
\cuniv(u_{1},\ldots,u_{n}).
\eeq
\eTheorem
\bProof
Let $p=\cuniv(u_1,\ldots,u_n)$.
Then, we have $a\And p\commutes b\And p$
for any $a,b\in L(u_1,\ldots,u_n)$, and hence 
there is a Boolean sublogic $\cB$ such that 
$L(u_1,\ldots,u_n)\And p\subseteq \cB$.
From Proposition \ref{th:L-restriction},
we have $L(u_1|_p,\ldots,u_n|_p)\subseteq \cB$.
From Proposition \ref{th:sublogic}, we have
$u_1|_p,\ldots,u_n|_p\in \VB$.
By the ZFC transfer principle of the Boolean-valued
universe \cite[Theorem 1.33]{Bel05}, we have
$\val{\ph(u_1|_p,\ldots,u_n|_p)}{}_{\cB}=1$. By the
$\De_0$-absoluteness principle, we have
$\val{\ph(u_1|_p,\ldots,u_n|_p)}=1$.
From Proposition \ref{th:V-restriction}, we have
$\val{\ph(u_1,\ldots,u_n)}\And p
=\val{\ph(u_1|_p,\ldots,u_n|_p)}\And p
=p$, and the assertion follows.
\eProof

\section*{Acknowledgements}
The author acknowledges the support of the JSPS KAKENHI, No. 26247016, No. 17K19970, 
and the support of the IRI-NU collaboration.
The author thanks the referee for calling his attention to the well-definedness of restrictions of quantum sets.

\end{document}